\documentclass[twocolumn,showpacs,preprintnumbers]{revtex4}
\usepackage{amsmath}
\usepackage{graphicx}
\usepackage{dcolumn}
\usepackage{bm}

\input{tcilatex}

\begin{document}

\preprint{APS/123-QED}
\title{Modelling of SARS for Hong Kong}
\author{Pengliang Shi}
\email{pengliangshi@yahoo.com}
\author{Michael Small}
\affiliation{Department of Electronic and Information Engineering, The Hong Kong
Polytechnic University, Hung Hom, Kowloon, Hong Kong}
\date{\today }

\begin{abstract}
A simplified susceptible-infected-recovered (SIR) epidemic model and a
small-world model are applied to analyse the spread and control of Severe
Acute Respiratory Syndrome (SARS) for Hong Kong in early 2003. From data
available in mid April 2003, we predict that SARS would be controlled by
June and nearly 1700 persons would be infected based on the SIR model. This
is consistent with the known data. A simple way to evaluate the development
and efficacy of control is described and shown to provide a useful measure
for the future evolution of an epidemic. This may contribute to improve
strategic response from the government. The evaluation process here is
universal and therefore applicable to many similar homogeneous epidemic
diseases within a fixed population. A novel model consisting of map systems
involving the Small-World network principle is also described. We find that
this model reproduces qualitative features of the random disease propagation
observed in the true data. Unlike traditional deterministic models,
scale-free phenomena are observed in the epidemic network. The numerical
simulations provide theoretical support for current strategies and achieve
more efficient control of some epidemic diseases, including SARS.
\end{abstract}

\pacs{89.75.Hc, 87.23.Ge}
\keywords{SARS, SIR, Small-World, epidemic}
\maketitle

During 2003 SARS killed 916 and infected 8422 globally\cite{WHO1}. In Hong
Kong (HK), one of the most severely affected regions, 1755 individuals were
infected and 299 died\cite{HK1}. SARS is caused by a coronavirus, which is
more dangerous and tenacious than the AIDS virus because of its strong
ability to survive in moist air and considerable potential to infect through
close personal contact\cite{Lawrence1,Ksiazek1,Drosten1,Peiris1}. Unlike
other well-known epidemic diseases, such as AIDS, SARS spreads quickly.
Although significant, its mortality rate is, fortunately, relatively low
(approximately 11\%)\cite{WHO1}. Researchers have decoded the genome of SARS
coronavirus and developed prompt diagnostic tests and some medicines, a
vaccine is still far from being developed and widely used\cite%
{Marra1,Rota1,Stadler1}. The danger of a recurrence of SARS remains.

Irrespective of pharmacological research, the epidemiology study of SARS
will help to prevent possible spreading of similar future contagions.
Generally, current epidemiological models are of two types. First, the
well-known Susceptible-Infected-Recovered (SIR) model proposed in 1927\cite%
{Kermack1,Daley1}. Second, the concept of Small-World (SW)\ networks\cite%
{Watts1}. Arousing a new wave of epidemiological research, the SW model has
made some progress recently\cite{Damian1,Holl1,Mode1}. Our work aims to
model SARS data for HK. Practical advice for a better control are drawn from
both the SIR and SW models. In particular, a generalized method to evaluate
control of an epidemic is promoted here based on the SIR model with fixed
population. Using this method, measuring the spread and control of various
epidemics among different\ countries becomes simple. Quick action in the
early stage is highlighted for both government and individuals to prevent
rapid propagation.

\section{Susceptible-Infected-Recovered model}

In the SIR model, the fixed population $N$ is divided into three distinct
groups: Susceptible $S$, Infected $I$ and Removed $R$. Those at risk of the
disease are susceptible, those that have it are infected and those that are
either quarantined, dead, or have acquired immunity are removed. The
following flow chart shows the basic procession of the SIR model \cite%
{Daley1}.%
\begin{equation}
\begin{array}{ccccc}
& rSI &  & aI &  \\ 
S & \rightarrow & I & \rightarrow & R%
\end{array}
\label{SIR1}
\end{equation}%
Here $r$ and $a$ are the infection coefficient and removal rate,
respectively. A discrete model was adapted by Gani from the original SIR
model through the coarse-graining process and was applied to successfully
predict outbreak of influenza epidemics in England and Wales\cite%
{Gani1,Spicer1}.

\begin{eqnarray}
S_{i+1} &=&S_{i}\left( 1-rI_{i}\right)  \notag \\
I_{i+1} &=&\left[ rS_{i}+\left( 1-a\right) \right] I_{i}  \label{SIR2} \\
R_{i+1} &=&R_{i}+aI_{i}  \notag
\end{eqnarray}%
During the epidemic process, $N=S_{i}+I_{i}+R_{i}$ is fixed.

\begin{figure}[h]
\centering\includegraphics[width=3.3in]{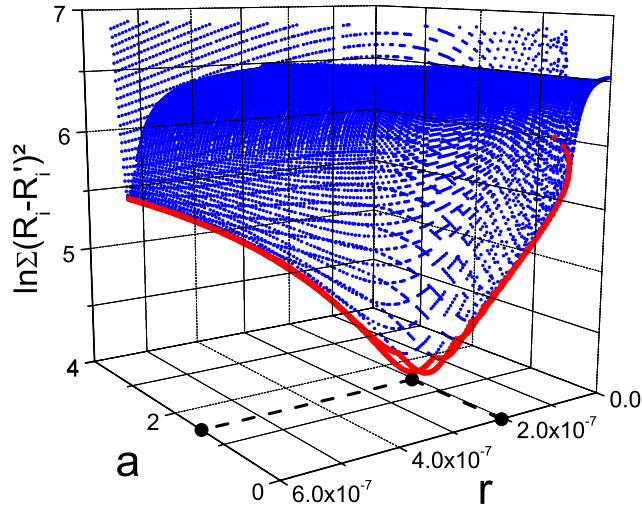}
\caption{The quantity $\sum_{i=1}^{15}\left( R_{i+1}-R_{i+1}^{\prime
}\right) ^{2}$ vs. $r$ and $a$ is plotted as dots for the first 15 days SARS
data for HK. The natural logarithm is applied. $R$ and $R^{\prime }$ are
original and simulated data, respectively. The thick red curve shows the
bottom of the sharp valley clearly. The lowest point of the valley is
according to the best fit: $r=2.05\times 10^{-7}$ and $a=1.444$.}
\label{graph1}
\end{figure}

\begin{figure}[h]
\centering\includegraphics[width=3.3in]{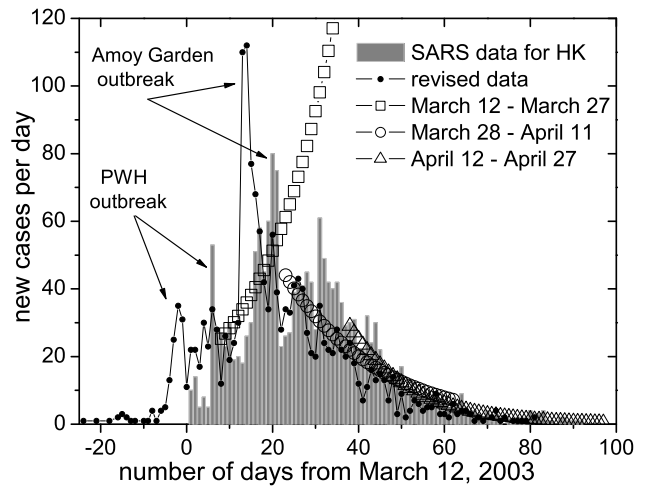}
\caption{SARS data announced daily by the Hong Kong Dept. of Heath, the
epidemic curve is plot as column graph. The curves of squares ($r=2.05\times
10^{-7}$ and $a=1.444$), circles ($r=1.55\times 10^{-7}$ and $a=1.172$) and
triangles ($r=1.09\times 10^{-7}$ and $a=0.868$) are prediction for the 3
stages described in the text. Each stage has 15 days. The curve of black
dots is for revised data.}
\label{graph2}
\end{figure}

\begin{figure}[h]
\centering\includegraphics[width=3.3in]{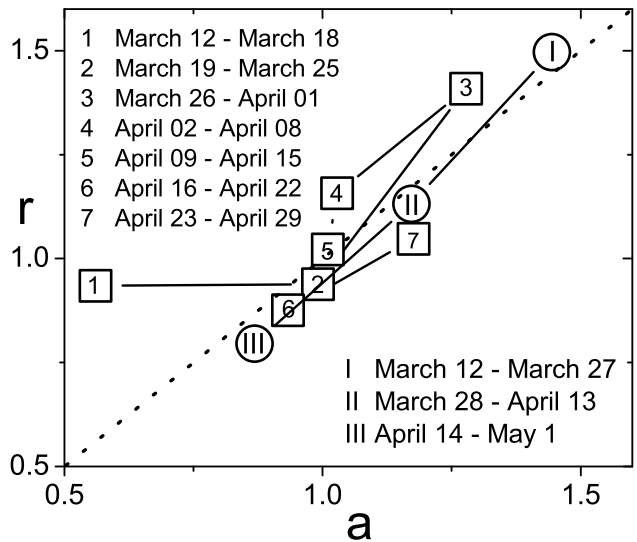}
\caption{The best fit\ parameters estimated week by week from SARS data of
HK are plotted as squares. The circles are for 3-stage analysis mentioned in
Fig. \protect\ref{graph2}. In this panel the line of dots $r=a$ is regarded
as \textquotedblleft alert line\textquotedblright\ or \textquotedblleft
critical line\textquotedblright . The parameters of $r$ and $a$ below it
indicate that the epidemic is controlled, above the line indicates
uncontrolled growth. The parameter $r$ is rescaled by $rN\rightarrow r$.}
\label{graph3}
\end{figure}

Initially, we examine the data for the first 15 days to estimate the
parameters $r$ and $a$ of SARS for HK. The only data are new cases (removal $%
R$) announced everyday by HK Dept. of Health from March 12, 2003 followed by
a revised version later\cite{HK1}. To avoid inadvertently using future
information we do not use the revised data at this stage. Population $%
N=7.3\times 10^{6}$; since $I+R\ll N$ it is reasonable to let $S_{i}=N$ in
right hand side of (\ref{SIR2}). $I_{1}=1$ and $R_{1}=0$ is set as the
initial condition. $I_{i}$ is replaced with $R_{i-1}$ whereas $I_{i}$ is
uncertain. This assumption implies the incubation period is only one day, in
spite of the fact that the true incubation period of the coronavirus\ is 2-7
days\cite{Rota1}. The parameters $r$ and $a$ are scanned for the best fit
for the stage. For every $\left( r,a\right) $, a sequence of $%
R_{i+1}^{\prime }$ is obtained by numerical simulations. An Euclidean norm
of $\sum_{i=1}^{15}\left( R_{i+1}-R_{i+1}^{\prime }\right) ^{2}$, which
indicates a distance between the true and simulated data, is applied to
measure the fit. In Fig. \ref{graph1} $\left( r,a\right) $ of the lowest
point is the best fit parameters for this stage and this value is used for
the following prediction. We get $r=2.05\times 10^{-7}$ and $a=1.444$. The
method is applied to get parameters of $a$ and $r$ in Figs. \ref{graph2} and %
\ref{graph3}.

The prediction is available for the trend based on parameters $r$ and $a$ of
this stage, the middle day (March 20, 2003) of the stage is applied as the
first day and $\overline{R}=\frac{1}{15}\sum_{i=1}^{15}R_{i+1}$ is assumed
as $\overline{I}$. A curve of squares is plotted in Fig. \ref{graph2} for
the first stage prediction. In the same way the best fit and prediction is
applied for the next two stages and plotted in Fig. \ref{graph2}. The best
fit is also processed weekly for detail discussion later. 

The curves for prediction clearly shows the 3 stages in Fig. \ref{graph2}.
The first stage (March 12 - 27, 2003) exhibits dangerous exponential growth.
It shows that the extremely infectious SARS coronavirus spread quickly in
the public with few protections during this early stage. More seriously, it
leads to a higher infection peak although in this stage the averaged number
of new cases $\overline{R}$ is below 30. This prediction gets confirmation
in the second stage (March 28 - April 11, 2003). The peak characterized by
the Amoy Garden outbreak comes earlier and higher. It indicates appearance
of a new transmission mode that differs from intimate contact route observed
in the first stage. We name the new transmission mode $explosive$ growth in
the contrast to transmission in the first stage which we refer to as $%
burning $ growth. Outbreaks at Prince Wales Hospital (PWH) (where SARS
patients received treatment) and Amoy Gardens (a high-density housing estate
in Hong Kong) represent infections of $burning$ and $explosive$ modes in the
full epidemic, respectively. The revised data provided by HK Dept. of Health
in Fig. \ref{graph2} present a clearer indication for these two transmission
modes. The outbreak of PWH sent the clear message that intimate contacts
with SARS patient led to infection\cite{Tomlinson1}. Detailed investigations
of the propagation at Amoy Garden suggested that faulty of sewerage pipes
allowed droplets containing the coronavirus to enter neighboring units
vertically in the building\cite{Riley1}. Furthermore, poor ventilation of
lifts and rat infestation where also suggested as possible modes of
contamination \cite{Normile1,Ng1,Helen1,Donnelly1}. However, control
measures bring the spread of the disease under control in the second stage.
Contrary to the increasing trend in the first stage, the prediction curve of
the 2nd stage (circle) declines. And it anticipates that new cases drop
below 10 before the 60th day (May 10, 2003). Also on April 12, 2003 we
predicted number of whole infection cases would reach 1700. Up to April 11,
2003 there were 32 deaths and 169 recoveries\cite{HK1}. We calculated the
mortality of SARS as the ratio of death to sum of both deathes and
recoveries, and it was 15.9\%. Therefore we predicted approximately 270
fatalities in total. In the third stage (April 12 - 27, 2003), triangle
curve in Fig. \ref{graph2} refines results and gives more accurate
prediction. This stage predicts that new cases per day drops to 5 before the
62th day, May 12, 2003. The travel warning for HK was cancelled by WHO
because HK had kept new cases below 5 for 10 days since May 15, 2003. And,
finally, we predict that the total cases reaches 1730 and nearly 287 deaths
(up to April 27, 2003 there were 668 recovered and 133, the mortality
increased to 16.6\%). These numbers are very close to the true data\cite{HK1}%
. Precisely the method drawn from the SIR model has been verified for
prediction for full epidemic. However, the accuracy is only possible by the
first dividing the epidemic into separate "stages".

This problem of determining to what extent an epidemic is under control is
of greater strategic significance. Information on the efficacy of epidemic
control will help determine whether to apply more control policies or not,
and balance cost and benefit from them.\ For each individual the same
question will also inform the degree to which precautions are taken: i.e.
wearing a surgical mask to prevent the spread (acquisition) of SARS.
Obviously a way to estimate control level is required. Actually this is a
difficult problem because of significant statistical fluctuations in the
data. A quite simple method to evaluate control efficiency is discussed
below.

In the SIR model (\ref{SIR2}), if $\Delta I=I_{i+1}-I_{i}\leq 0$ a disease
is regards as being controlled as new cases will decrease. This inequality
leads to a control criterion for some diseases in epidemiology research.
Applying the approximation of $S_{i}=N$ we get a threshold $Nr\leq a$ from (%
\ref{SIR2}). We rescale $rN\rightarrow r$ ($r$ is called infection rate in
place of infection coefficient now) and then get the threshold that is free
from population $N$.%
\begin{equation}
r\leq a  \label{SIR3}
\end{equation}%
This indicates that the removal rate $a$ exceeds the infection rate $r$. In
Fig. \ref{graph3} the circles show the 3 stages evaluations with dash line
of $r=a$. And the parameters $r$ and $a$ of SARS\ data for HK are estimated
weekly as squares also in Fig. \ref{graph3}. The line of $r=a$ is regarded
as the critical line since number of infected cases increases when $\left(
r,a\right) $ passes through it from below. It is possible to apply the
diagram of $r$ and $a$ to compare control level for different countries and
areas even for different diseases. This provides organizations like WHO with
a simple and standard method to supervise infection\ level of any disease.
The limitation of the method comes from the assumptions of the SIR model.
More accurate models may provide better estimate of epidemic state and
future behaviours.

In summary, a discrete SIR model gives good predictions for epidemic of SARS
for HK. Two distinct methods are described for the disease propagation
dynamics. Particularly the $explosive$ mode is much more hazardous than the $%
burning$ mode. We have introduced a simple method to evaluate control
levels. The method is generic and can be widely applied to various
epidemiological data.

\section{The small-world model}

Contrary to the long established SIR model, epidemiology research using
Small-World (SW) network models is young and growing area. The concept of SW
was imported from the study of social network into nature sciences in 1998%
\cite{Watts1}. However, it provides a novel insight for networks and\
arouses a lot of explorations in the brain, social networks and the Internet.%
\cite{Laughlin1,Albert1,Ebel1,Liljeros1}. Some SW networks also exhibit a
scale-free (power law) distribution: a node in this network has probability $%
P(k)\sim k^{-\gamma }$ to connect $k$ nodes, $\gamma $ is one basic
character for the system\cite{Albert1,Newman1,Ebel1,Liljeros1}. Most
researches of \textquotedblleft virus\textquotedblright\ spreading with the
SW model, concern the propagation of computer viruses on the Internet,
however, a few studies have been published relating to epidemiology\cite%
{Ebel1,Liljeros1,Strogatz1,Kuperman1,Huerta1}. The dynamics of SW models
enriches our realization of epidemic and possibly provides better control
policies\cite{Miramontes1}. From the first SARS patient to the last one, an
epidemic chain is embedded on the scale free SW network of social contacts.
It is of great importance to discover the underlying structure of the
epidemic network because a successful quarantine of all possible candidates
for infection will lead to a rapid termination of the epidemic. In HK
e-SARS, an electronic database to capture on-line and in real time clinical
and administrative details of all SARS patients, provided invaluable
quarantine information by tracing contacts\cite{HK1,Brower1}. Unfortunately
a full epidemic network of SARS for HK is still unavailable. Because data
representing the underlying network structure is currently not available, we
have no choice other than numerical simulations. Therefore, our analysis of
the SW model is largely theoretical. The only confirmation of our model we
can offer is that the data appear to be realistic and exhibit the same
features as the true epidemic data.

To simulate an epidemic chain, a simple model of social contacts is proposed
. The model is established on a grid network weaved by $m$ parallel and $m$
vertical lines. Every node in the network represents a person. We set $%
m=2700 $ with population $N=m^{2}=7.29\times 10^{6}$. All nodes are
initiated with a value of 0 (named $good$ nodes). Every node has 2, 3 or 4
nearest neighbors as short range contacts for corner, edge and center,
respectively. For every node there are two long range contacts with 2 other
nodes randomly selected in the whole system everyday. These linkages model
the social contact between individuals (i.e. social contacts that are
sufficiently intimate to bring individuals at risk of spreading the
disease). One random node of the system is set to 1 (called the $bad$ node),
through its short and long range contacts, the value of the nodes linked
with it turns into 1 according to probability of $p_{1}$ and $p_{2}$,
respectively. An infection happens if a node changes its value from 0 to 1.
This change is irreversible. During the full simulation process, the bad
node is not removed from the whole system. We make this assumption because
the number of deaths is small in comparision of the population, and there is
no absolute quarantine---even the SARS patient in hospital can affect the
medical workers. Moreover, the treatment period for SARS is relatively
length, and during this time infected indviduals are highly infectious. To
reflect the true variation in control strategy and individual behaviour, the
control parameters $p_{1,2}$ ($0\leq p_{1,2}\leq 1$) vary with time.

\begin{figure}[h]
\centering\includegraphics[width=3.3in]{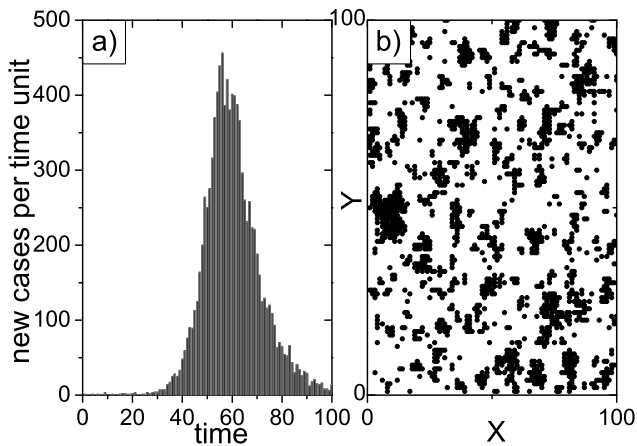}
\caption{A plague is simulated in a geographical map with $100\times 100$
nodes with fixed infection probability $p_{1,2}=0.005$ which eventually
leads to a full infection. a). The full epidemic curve of the simulated
epidemic. b). While time is 45, the infection cases clusteringly scatter in
the map. The infection happens in full scale map because\ of random
connections. The clusters show effect from the nearest neighbour
connections. }
\label{graph4}
\end{figure}

\begin{figure}[h]
\centering\includegraphics[width=3.3in]{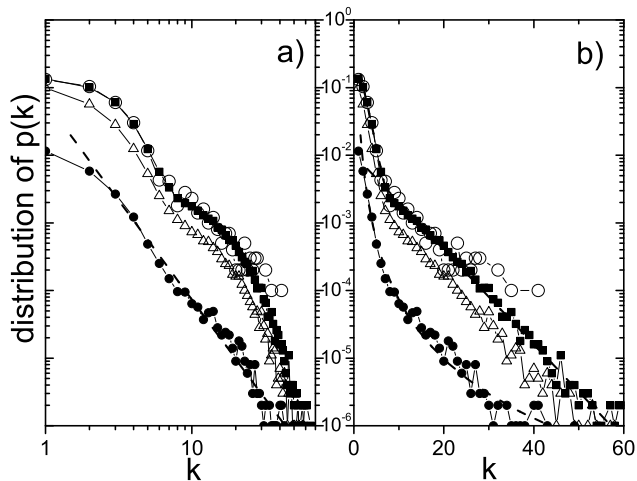}
\caption{The scaling behaviour for a full infection epidemic network is
plotted in log-log and log-linear coordinates in a) and b), respectively.
The circle is for a full infection epidemic network simulated in a
geographical map with $100\times 100$ nodes. The curves are for one with $%
1000\times 1000$ nodes. The curves of dots, triangles and squares are for
the epidemic networks of 1\% , 30\% and 100\% for full infection process.
The two dash lines are piecewise linear fits for the full infection process
with $1000\times 1000$ nodes. The ratio is $-0.31$ and $-0.07$ with $%
R=-0.997 $ and $-0.994$, respectively. It clearly shows the curves of full
epidemic has two piecewise linear parts in log-linear graph. In the early
stage of a full infection, the scaling behaviour of 1\% infected (dots) has
a part which might be regards as a log-log linear. The ratio of it is $-2.89$
with $R=-0.983$. In other words, it has scale-free part with $\protect\gamma %
=-2.89 $.}
\label{graph5}
\end{figure}

In same model it would be interesting to compare with the complete infection
epidemic. A small system with $100\times 100$ nodes is chosen. A fixed
probability\ $p_{1}=p_{2}=0.05$ leads to eventual infection of the entire
population since all nodes are linked and the infected are not removed. The
epidemic curve for this process is plotted in Fig. \ref{graph4} a) and is
typical of many plagues. In Fig. \ref{graph4} b) various sizes of clusters
with infected nodes (black dots) scatter over the geographical map. It
contains all infection facts in first 45 days. Comparing to the true SARS
infection distribution in HK only slight similarity is observed. The short
and long range linkages give good infection dynamics as we expected. The
epidemic chain is easily drawn by recording infection fact from the seed\ to
last patient. However, during simulations there is a problem: for a $good$
node linked by more than one $bad$ nodes, which $bad$ node infects it? A
widely accepted preferential attachment of \textit{rich-get-richer} is a
good answer\cite{Albert1,Newman1,Watts2}. A linear preference function is
applied here. With presence of both growth and preferential attachment, it
is general to ask whether the chain is a scale-invariant SW network. We plot
the distribution in Fig. \ref{graph5} a) and b) in log-log and log-linear
coordinates, respectively. The hollow circle is for the system with $%
100\times 100$ nodes. The scaling behaviour of it looks more like a
piecewise linear (i.e. bilinear) in b) rather than a power law in a). To
confirm this we tried a larger system with $1000\times 1000$ nodes and the
same\ fixed $p_{1,2}$ in the rest curves in Figs. \ref{graph5}. The solid
square curve is distribution of the whole epidemic network. The linear fit
for the solid square gives a correlation coefficient $R$ of $-0.956$ in Fig. %
\ref{graph5} a). In b) piecewise linear fits have $R$ of $-0.997$ and $%
-0.994 $. These provide positive support for the original model. The ratio
of the cases of first 30\% and 1\% of whole process for this bigger system
are plotted as hollow triangle and solid circle curves in Figs. \ref{graph5}%
, respectively. The case of 1\%, early stage of full infection epidemic,
suggests a better fit for scale-free than the other cases since it has
correlation coefficients $R$ of $-0.983$ and ratio $\gamma =-2.89$ for
linear fit in a). For the solid circle curve piecewise linear fits give $R$
of $-0.995$ and $-0.977$ in Fig. \ref{graph5} b). So, what scaling behaviour
is true during full infection process? With absence of rigorous proof this
problem is hard to answer correctly in simulations. We may only draw
conclusions based on which simulations most closely matches the qualitative
features of the observed data.

\begin{figure}[h]
\centering\includegraphics[width=3.3in]{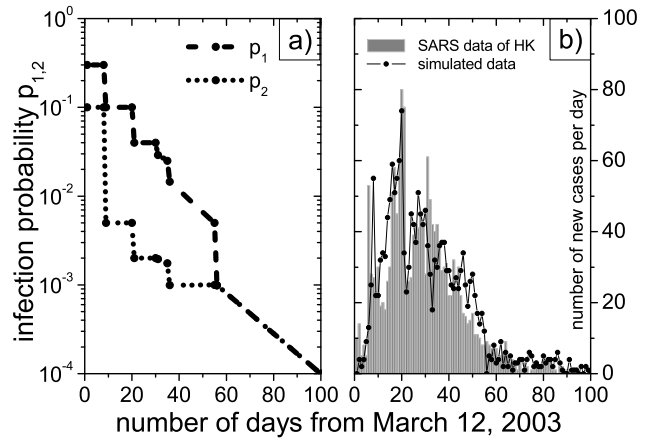}
\caption{Control parameters and epidemic curve is plotted for the model with 
$2700\times 2700$ nodes. a) The short and long range linkages infection
probability $p_{1,2}$ generally declines exponentially. b) The simulated
epidemic curve (black dots) in the model is plotted with the original SARS
data (grey column) for HK.}
\label{graph6}
\end{figure}

Let's return to the system with $2700\times 2700$ nodes for modelling SARS
for HK. The probability $p_{1,2}$ is fitted to the true epidemic data.
However, it is fruitless to obtain an exact coincidence between the
simulated results and the true data as the model evolution is highly random
(moreover this would result in overfitting). The control parameters $p_{1,2}$
(dots and dashes curves)\ and a simulated epidemic curve (black dots) with
column diagram of SARS for HK is plotted in Figs. \ref{graph6} a) and b),
respectively. For the simulated data, the total number of cases is 1830 that
has a 4.3\% deviation from the true data of 1755. Contrary to the above full
infection with fixed parameters, $p_{1,2}$ are believed to drop
exponentially and lead to small part infection epidemic without quarantine
or removal. In any case, the high probability of infection in the early
stage is indicative of the critical ability of the SARS coronavirus to
attack an individual without protection. If SARS returns, the same high
initial infection level is likely to occur. The only hope to avoid a repeat
of the SARS crisis of 2003 is to shorten the high infection stage by quick
identification, wide protection and sufficient quarantining. In other words,
the best time to eliminate possible epidemic is the moment that the first
patient surfaces. Any delay may lead to a worsening crisis. The long range
infections in the model and the world also imply an efficient mechanism to
respond rapidly to any infectious disease is required to establish global
control.

\begin{figure}[h]
\centering\includegraphics[width=3.3in]{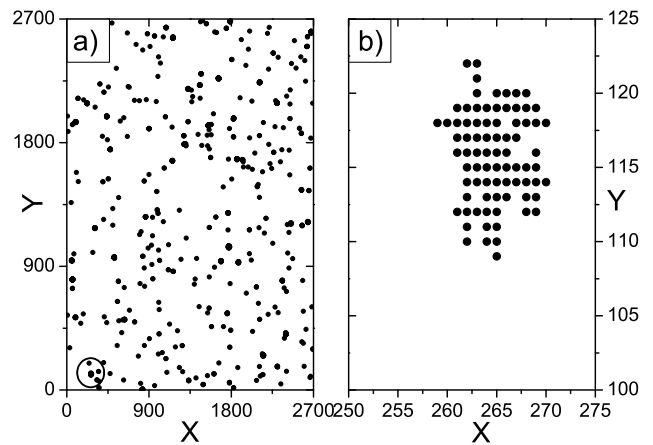}
\caption{The distribution of infected nodes in simulations in two
dimensional map. a). the whole geographical map; b). an amplified window of
a cluster marked in circle of a).}
\label{graph7}
\end{figure}

\begin{figure}[h]
\centering\includegraphics[width=3.3in]{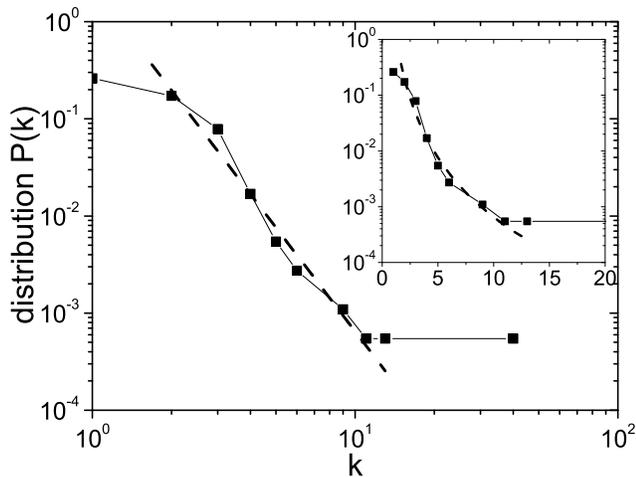}
\caption{The scaling behaviour for the simulated epidemic network. In
log-log coordinates, the scaling $P(k)\sim k^{-\protect\gamma }$, $\protect%
\gamma =-3.55$, where the linear fit exhibits a correlation coefficient $%
R=-0.989$.}
\label{graph8}
\end{figure}

Data on the geographical distribution of SARS cases in HK is much easier to
collect than the full epidemic chain. Numerical simulations provide both
simply. The full geographical map marked with all infected nodes (black
dots) and an amplified window of a cluster is plotted in Figs. \ref{graph7}
a) and b), respectively. Similarity is expected and verified in cluster
patterns of Figs. \ref{graph4} b). The scaling behaviour of the epidemic
chain is plotted in Figs. \ref{graph8}. The curve in log-log diagram
exhibits a power-law coefficient of $\gamma =-3.55$ and gives the linear fit
correlation coefficient $R=-0.989$. The piecewise linear fit for the
log-linear case gives correlation coefficients $R$ of $-0.987$ and $-0.959$,
respectively. 

For a SW network often few vertexes play more important roles than others%
\cite{Watts2}. SARS super-spreaders found in HK, Singapore and China are
consistent with this\cite{Riley1,Leo1}. Data for the early spreading of SARS
in Singapore\cite{Leo1} show definite SW structure with a small number of
nodes with a large number of links. The average number of links per node
also shows a scale-free structure \cite{Leo1}, but the available data is
extremely limited (the linear scaling can only be estimated from three
observations).

This character is also verified in our model. The first few nodes have a
high chance of infecting a large number of individuals. In Fig. \ref{graph8}%
, a single node has 40 links.\ Clearly the index node has many long range
linkages. It has been suggested that travelling in crowded public places
(train, hospital, even an elevator) without suitable precautions can cause
an ordinary SARS patient to infect a significant number of others. Again,
this is an indication that to increase an individual's (especially a
probable SARS patient's) personal protection is key to rapidly control an
epidemic. Actually, in our model, if duration of the early stage with high
probability $p_{1,2}$ is reduced to less than 10, the infection scale
decreases sharply.

An engrossing phenomenon is the points of inflection in curves in log-linear
diagrams of Figs. \ref{graph8} and Figs. \ref{graph5} b). All are located
near linkages number of 6-7. On average a nodes has about 6 contacts (2-4
short plus 2 long range) every day, although in whole process there is no
limitation for linkages. For a growing random network, a general problem
always exists among its scaling behaviour, preferential attachment and
dynamics, even embedding a geographical map\cite%
{Cohen1,Warren1,Krapivsky1,Dorogovtsev1,Moore1}. More work is required to
address this issue.

In conclusion a SW epidemic network is simulated to model SARS spreading in
HK. A comparison of the simulations with full infection data is presented.
Our discussion of the infection probability and occurrence of super
spreaders lead to the obvious conclusion: rapid response of an individual
and government is a key to eliminating an epidemic with limited impact and
at minimal cost.

\begin{acknowledgments}
This research is supported by Hong Kong University Grants Council CERG
number B-Q709.
\end{acknowledgments}

\end{document}